\documentclass[conference,times,mathptm,psfig]{IEEEtran}
\IEEEoverridecommandlockouts

\usepackage{textpos}
\newcommand\copyrighttext{%
	\scriptsize \textcopyright 2022 IEEE. Personal use of this material is permitted. Permission from IEEE must be obtained for all other uses, in any current or future media including reprinting/republishing this material for advertising or promotional purposes, creating new collective works, for resale or redistribution to servers or lists, or reuse of any copyrighted component of this work in other works. }
\newcommand\copyrightnotice{%
	\begin{tikzpicture}[remember picture,overlay]
		\node[anchor=north,yshift=-5pt] at (current page.north) {\fbox{\parbox{\dimexpr\textwidth-\fboxsep-\fboxrule\relax}{\copyrighttext}}};
	\end{tikzpicture}%
}

\usepackage{bbm}
\usepackage{amsmath}
\usepackage{bm}
\usepackage{mdframed}
\usepackage{amsthm}

\usepackage[utf8]{inputenc}
\usepackage{graphicx}
\usepackage[colorinlistoftodos]{todonotes}
\usepackage{amssymb}
\usepackage{longtable}
\usepackage[colorlinks=true, allcolors=black]{hyperref}

\usepackage[noadjust]{cite}
\allowdisplaybreaks
\usepackage{lineno}
\usepackage{lipsum}
\usepackage{multicol}
\usepackage{subcaption}

\usepackage{multirow}
\usepackage{array}

\usepackage{xcolor}
\usepackage[vlined,linesnumbered,ruled,resetcount]{algorithm2e}
\SetArgSty{textnormal}

\usepackage{bbm}

\SetKwInput{KwInput}{Input}                
\SetKwInput{KwOutput}{Output}              
\SetKwInput{KwInitialize}{Initialize}                
\SetKwInput{myproc}{Procedure}{}{}

\renewcommand{\baselinestretch}{0.93}

\title{Learning-Based Orchestration for Dynamic Functional Split and Resource Allocation in vRANs}

\author{\IEEEauthorblockN{Fahri Wisnu Murti$^\dagger$, Samad Ali$^\dagger$, George Iosifidis$^*$, Matti Latva-aho$^\dagger$}\\
\vspace{-4mm}
\IEEEauthorblockA{
	$^\dagger$Centre for Wireless Communications, University of Oulu, Finland\\
	$^*$Delft University of Technology, Netherlands
}
\thanks{{This research has been supported by the Academy of Finland, 6G Flagship program under Grant 346208.}}%
}

\begin{document}

\maketitle

\begin{abstract}
One of the key benefits of virtualized radio access networks (vRANs) is network management flexibility. However, this versatility raises previously-unseen network management challenges. In this paper, a learning-based zero-touch vRAN orchestration framework (LOFV) is proposed to jointly select the functional splits and allocate the virtualized resources to minimize the long-term management cost. First, testbed measurements of the behaviour between the users' demand and the virtualized resource utilization are collected using a centralized RAN system. The collected data reveals that there are non-linear and non-monotonic relationships between demand and resource utilization. Then, a comprehensive cost model is proposed that takes resource overprovisioning, declined demand, instantiation and reconfiguration into account. Moreover, the proposed cost model also captures different routing and computing costs for each split. Motivated by our measurement insights and cost model, LOFV is developed using a model-free reinforcement learning paradigm. The proposed solution is constructed from a combination of deep Q-learning and a regression-based neural network that maps the network state and users' demand into split and resource control decisions. Our numerical evaluations show that LOFV can offer cost savings by up to 69\% of the optimal static policy and 45\% of the optimal fully dynamic policy.
\end{abstract}

\IEEEpeerreviewmaketitle
\vspace{-2mm}
\copyrightnotice
\vspace{-2mm}
\section{Introduction}
\vspace{-2mm}
Virtualizing the radio access network (vRAN) is one of the most promising technologies for accommodating the increased service demands with diverse requirements at a reasonable cost \cite{nokia5g}. Its latest development enables the base station (BS) functions to turn into virtualized components that can be executed across cloud platforms. This paradigm brings unprecedented flexibility to RAN operations, mitigates vendor lock-in, offers fast deployment and potentially reduces operational expenses \cite{openvran_nec}.  
Therefore, it is not surprising that many standardization bodies adopt virtualization for Next-Generation RANs such as Open RANs \cite{oran_architecture} and 5G+ RANs \cite{3gpp_rel16}.

In vRANs, the virtualized BS (vBS) functions can be disaggregated then hosted at virtualized distributed units (vDUs) and central units (vCUs) through \textit{functional split}. The network operators can flexibly deploy their vBS functions over vCUs and vDUs based on their resource availability and the network load, facilitating cost-efficient and high-performance RANs operation \cite{vran_optimal_murti2}. However, this flexibility also carries non-trivial decisions on splitting the functions and allocating the virtualized computing resources\footnote{It is common in Network Function Virtualization (NFV) that Virtualized Network Functions (VNFs) needs CPU, memory, I/O, and GPU for their virtualized computing resources. The operators typically use these parameters to calculate their billing units to charge the amount of monetary cost.} to implement these functions. 
Each split induces different fronthaul/backhaul (xHaul) load and virtualized resources. The suitability of each split also depends on the network properties (xHaul capacity, computing capacity, etc.) and might change abruptly over time due to the varying user needs. 
%
%
Therefore, it is not only important to design the splits and resource allocation in vRANs carefully, but also to update these decisions and reconfigure the system in order to adapt to varying conditions (resource availability and user needs). 
Otherwise, we risk inducing high operating expenditures and performance degradation. 

On the other hand, orchestrating the dynamic split selection and virtualized resource allocation is also a challenging problem as the decisions take place before the actual users' traffic is observed. Hence, there is a risk for resource \textit{overprovisioning} (e.g., the allocated resources are higher than the actual usages resulting in waste resources) and \textit{underprovisioning} (e.g., insufficient allocated resources resulting in declined users' demand). Meanwhile, reconfiguring the splits and virtualized resources at runtime can induce additional costs, potentially disrupts network operations during the migration of the virtual machines (VMs) \cite{adaptive_vran_alba}, and \textit{is therefore not always beneficial}.
Conversely, deploying a static policy can not unleash the potential flexibility of the vRAN system. Therefore, it is necessary to manage and reconfigure the splits and virtualized resources in an intelligent manner.

%
The authors of \cite{adaptive_vran_alba} have experimentally analyzed the migration activity of VMs in vRANs and have successfully validated the possibility of deploying an adaptive functional split practically, albeit have not discussed designing the cost-efficient split reconfiguration policies. Recent work in \cite{dynamic_split_alba} has proposed a framework for the split that dynamically adapts its configuration at runtime to maximize the users throughput. Similar works have proposed flexible split selection 
to minimize the inter-cell interference and fronthaul utilization \cite{flex5g} and the network cost \cite{vran_optimal_murti2}. However, these works assume complete knowledge models of the split performance and resource utilization.
We argue that such approaches can be inaccurate in practice as vRANs softwarization is deployed together with other workloads in the diverse cloud platform, which are hardly precise in predicting its resources and behaviour. 

Using Machine Learning (ML) techniques for tackling resource allocation problems in wireless networks is becoming increasingly popular \cite{ali20206g}. 
The authors in \cite{vranai_journal} have proposed a learning framework that successfully manages the interplay between computing and radio resources. It models the problem as a contextual bandit, then utilizes an actor-critic neural network structure and a classifier to map contexts into resource control decisions. The follow-up work \cite{bayes_vran} has studied an energy-aware resource orchestration that uses Bayesian online learning to balance performance and energy consumption. The authors in \cite{concordia} have proposed an ML-based predictor that learns to share the unutilized CPU resources with the other workloads such as video analytics. 
Recent works have brought the importance of ML-based optimization \cite{murti_iccw_cdrs_vran} and dynamic selection \cite{rl_oran} of functional splits, albeit not discussing virtualized resource management. Although the mentioned works have addressed complex vRANs problems, they still did not answer how to design a framework that intelligently decides the splits and allocates the virtualized resources. Moreover, they also did not consider the effects of resource reconfiguration in their vRAN problems.


\textbf{ Contributions.} We firstly use a vBS prototype implementing the srsRAN platform \cite{srslte} to collect measurements regarding the behaviour between the users' demand and the virtualized resource utilization (details in Sec. \ref{sec:simulation}). Our findings suggest that this behaviour varies with the demand and the platform resources and, importantly, is \textit{non-linear} and \textit{non-monotonic}; hence it is hard to model the underlying system precisely. 
Inspired from network slicing \cite{aztec}, we also propose a new cost model accounting for resource overprovisioning, declined service demand, reconfiguration and instantiation, representing the virtualized resource management in vRANs.  Besides, depending on the splits, the transferred load between vCU, vDU and the radio unit (RU) induces a different cost for reserving the xHaul link bandwidth. 
%

Our goal is to develop a learning-based zero-touch orchestration framework (LOFV) that intelligently selects the splits and allocates the virtualized resources to minimize the long-term management cost while serving the users' demand. We model the vRAN operation as a time-slotted system, where each slot has an arbitrary incoming users' demand and network state conditions. At every time stage, LOFV decides whether to preserve the previous network settings or reconfigure them by reselecting the splits and reallocating virtualized resources. We formulate this sequential decision-making problem as Markov Decision Process (MDP).
%
%
In our solution, LOFV is tailored from a model-free reinforcement learning paradigm that does not make particular assumptions about the underlying system and state transition. It is constructed from two functional blocks based on neural network structures: \textit{i)} the functional split orchestration and \textit{ii)} virtualized resource orchestration. 
The functional split orchestration is a function that maps the input state into the selected network setting and deployed split. It is constructed from a deep Q-network (DQN) and target network \cite{dqn_mnih1}, and utilizes Q-learning \cite{rl_sutton} as the learning step. The virtualized resource orchestration is a regression-based neural network that maps the input of split selection and users' demand into resources at vDU and vCU. This function utilizes $\alpha$-OMC loss function \cite{deepcog} that considers the penalty fee incurred from prediction errors of virtualized resources.  

We conduct a battery of tests using our collected measurements from container-based virtualization of srsRAN \cite{srslte}. We evaluate our vBS measurements, learning convergence of LOFV and long-term accumulated cost during the online stages. Our evaluations show that the cost-saving of LOFV can be as high as 69\% of the optimal static policy (STAO)\footnote{STAO is the optimal static policy that selects the single best split and resource allocation based on the peak traffic. Consequently, STAO only incurs overprovisioning and xHaul costs. We use STAO to normalize all of our monetary cost evaluations. } and 45\% of the optimal fully dynamic policy (DYNO)\footnote{It knows an oracle of users' demand and resource availability. It uses a fully dynamic policy by always reconfiguring the vRAN settings at every time stage to obtain the current optimal, i.e., the current best split and resources.}.

The rest of this paper is organized as follows. The background, model and trade-offs are presented in Sec. \ref{sec:model}. The problem formulation and LOFV framework are discussed in Sec. \ref{sec:algo}. Our detailed experiments and results are in Sec. \ref{sec:simulation} and the conclusion is in Sec. \ref{sec:conclusion}.  



\vspace{-2.5mm}
\section{Model and Trade-offs} \label{sec:model}

\vspace{-2.5mm}
\subsection{Background}
\vspace{-1mm}
Our model refers to the O-RAN compliant system model \cite{oran_architecture}.
We consider a vBS comprising a vCU and vDU connected to RU, corresponds to 4G eNodeB or 5G gNodeB. The vDU is typically hosted at a far-edge cloud and vCU is at an edge cloud. Let suppose $f_0$ is a function that encapsulates RF. Then, we denote $f_1$, $f_2$ and $f_3$ for the respective functions of Layer 1 (PHY), Layer 2 (MAC, RLC) and Layer 3 (PDCP, RRC, GTP). These functions can be deployed at the vDU and vCU (except $f_0$) following a chain: $f_0 \!\!\rightarrow\!\! f_1 \!\!\rightarrow\!\! f_2 \!\!\rightarrow\!\! f_3$. We consider four split options that have been well standardized \cite{3gpp_rel16, smallcell} and experimentally validated as a prototype \cite{adaptive_vran_alba}. \textbf{Split 1 (S1):} All functions are at vDU except $f_0$ is at RU (a fully distributed-RAN). \textbf{Split 2 (S2):} $f_3$ is deployed at vCU, but $f_1$ and $f_2$ are at vDU. \textbf{Split 3 (S3):} $f_2$ and the higher layer are at vCU, while $f_1$ is at vDU. \textbf{Split 4 (S4):} All functions are at vCU except $f_0$ (a fully centralized vRAN). Hence, we define the respective split $ i \in \mathcal{I} =  \{1,2,3,4\} $. 
%
%
%
LOFV is to be executed from the Learning Agent (LA) inside Non-Real-Time (Non-RT) RAN Intelligent Controller (RIC). 

\begin{figure}[t!]
	\centering
	\begin{subfigure}[c]{0.235\textwidth}
		\includegraphics[width=\textwidth]{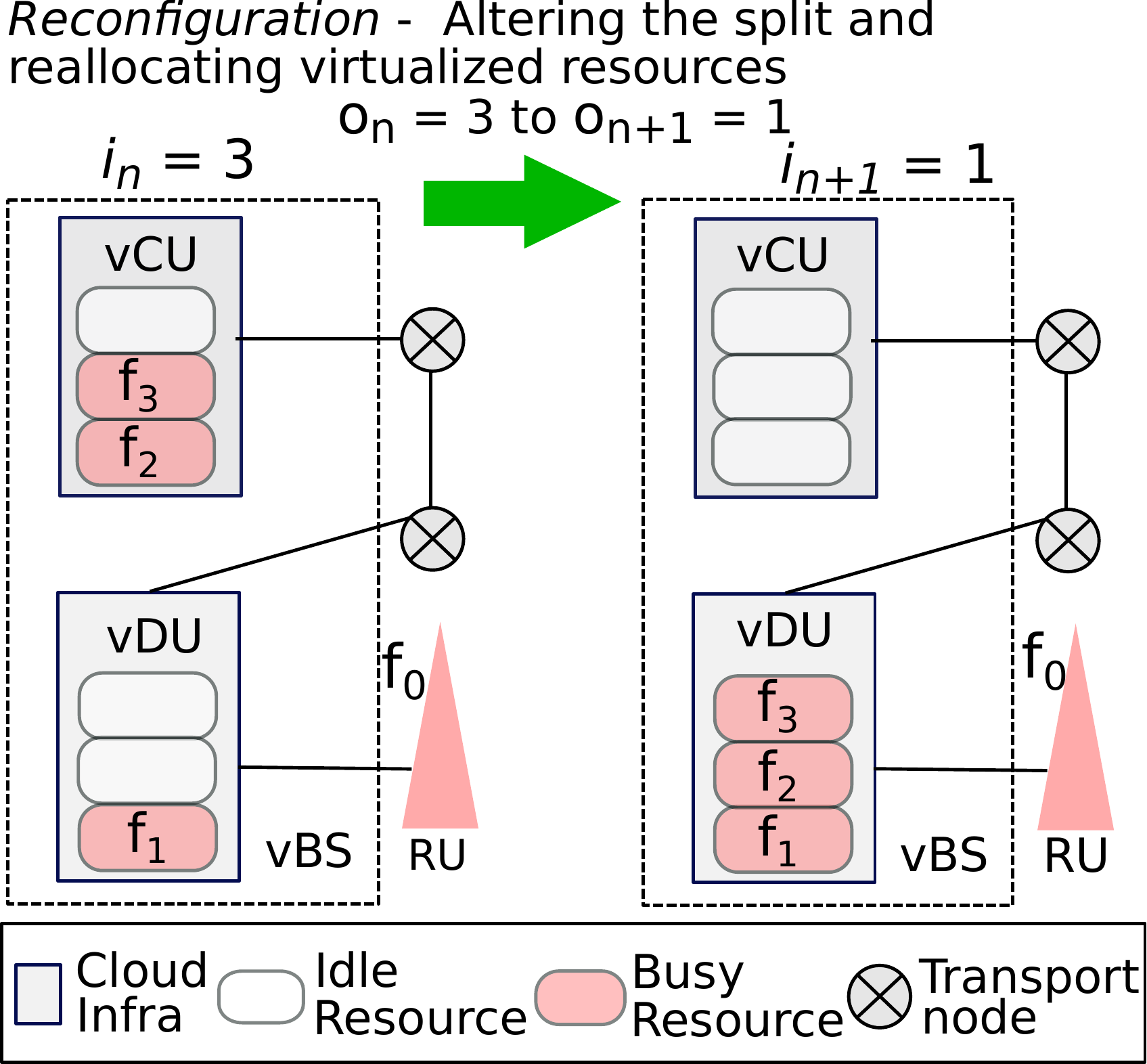}   
		\caption{}
		\label{fig:migration1}
	\end{subfigure}
	\hfill
	\begin{subfigure}[c]{0.235\textwidth}
		\includegraphics[width=\textwidth]{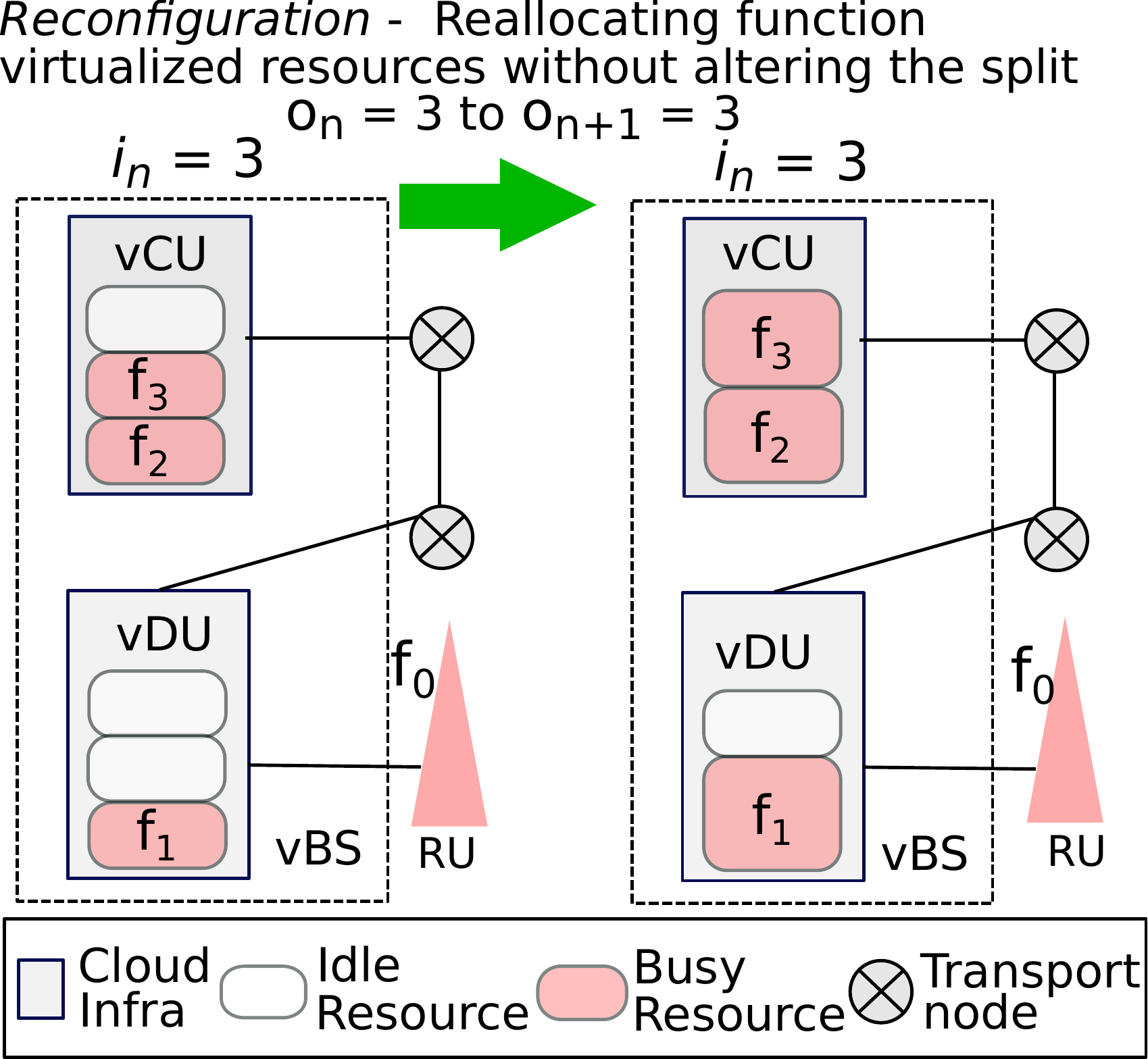}   
		\caption{}
		\label{fig:migration2}
	\end{subfigure}
	\hfill
	\begin{subfigure}[c]{0.235\textwidth}
		\includegraphics[width=\textwidth]{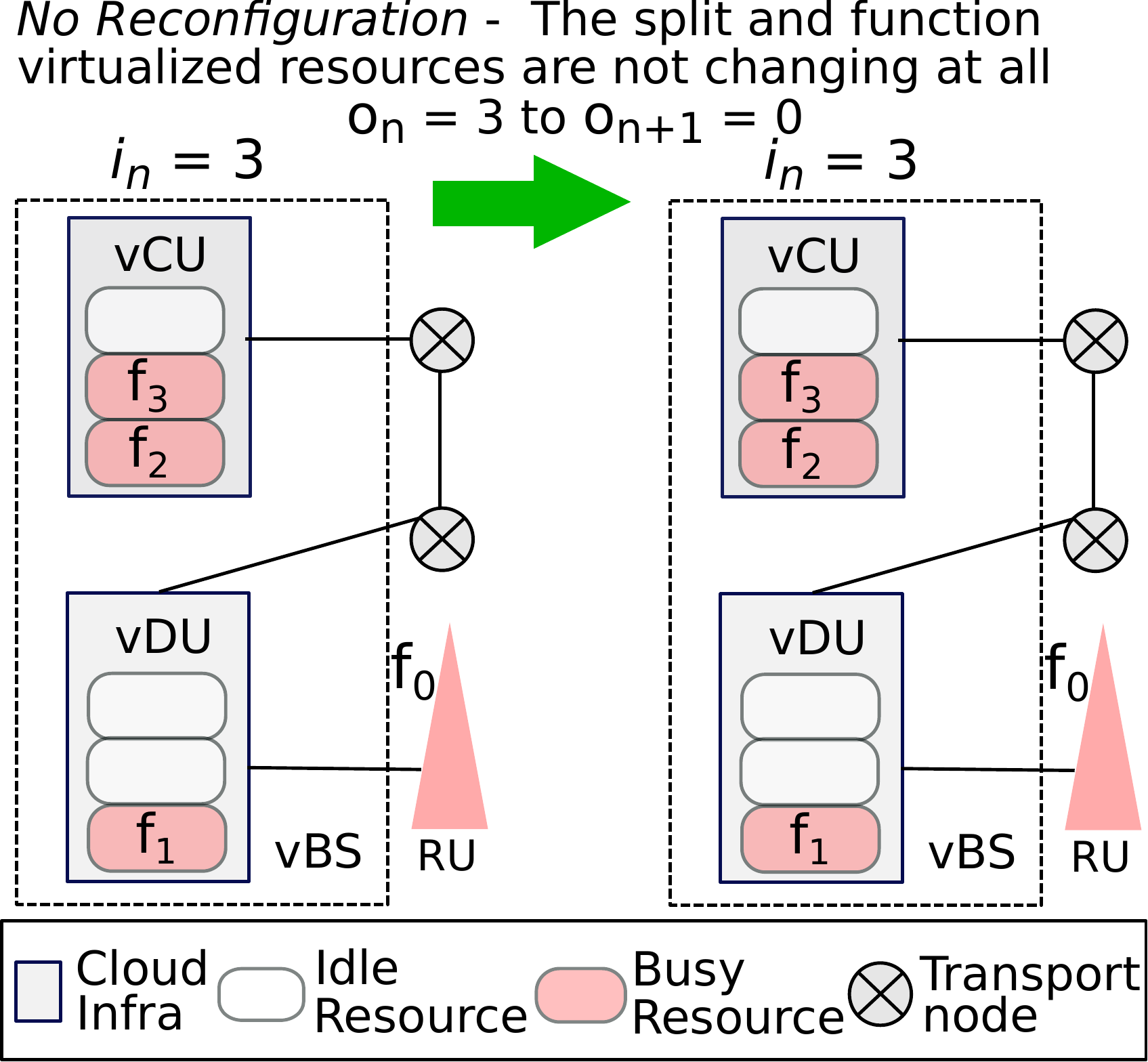}   
		\caption{}
		\label{fig:migration3}
	\end{subfigure}
	\hfill
	\begin{subfigure}[c]{0.235\textwidth}
		\includegraphics[width=\textwidth]{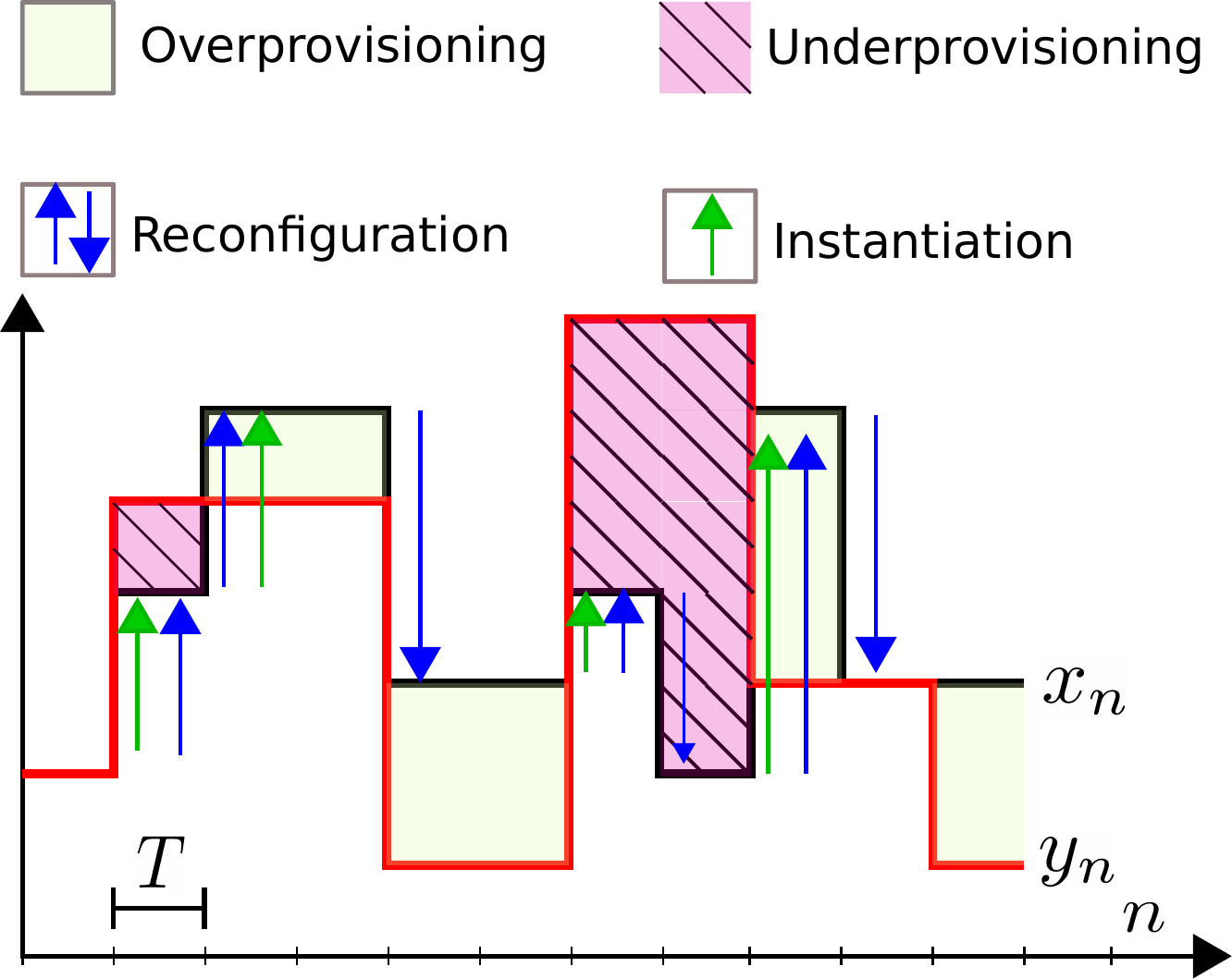}   
		\caption{}
		\label{fig:orchestration}
	\end{subfigure}
	\vspace{-2mm}
	\caption{\small \textbf{Virtualized Resource Management.} Reconfiguration activity occurs when the operators decide to: \textbf{(a)} alter the deployed split and reallocate the virtualized resources; \textbf{(b)} only reallocate the resources but keep the deployed split. Otherwise, \textbf{(c)} the operators may not want to reconfigure their resources and the deployed split at all. \textbf{(d)} The virtualized resource behaviour. }
	\vspace{-5mm}
\end{figure}

\vspace{-3mm}
\subsection{Virtualized Resource Management}
\vspace{-1.5mm}

%
Let $\lambda_n$ (Mbps) denote the incoming traffic demand from multiple users at time stage $n$ which has an interval of duration $T$. At the beginning of each interval, the operators need to decide network configuration setting $o_n \in \mathcal{O} = 0 \cup \mathcal{I}$, select split $i \in \mathcal{I}$, and allocate function virtualized resources at the vDU $x_n \in \mathbb{R}$ and the vCU $\hat{x}_n \in \mathbb{R}$. The option \textbf{0} ($o_n \!=\! 0$) is dedicated for \textit{no reconfiguration}\footnote{The last network settings are preserved; consequently, the selected split and virtualized resources at the current sequence same as the previous one: $o_n \!=\! 0 \!\iff\! \!(i_n \!=\! i_{n-1} \land x_n \!=\! x_{n-1} \land \hat{x}_n \!=\! \hat{x}_{n-1} )$. } (Fig. \ref{fig:migration3}). Then, a \textit{reconfiguration}\footnote{The operators may decide to alter the deployed split, but it will change the incurred resources at vDU and vCU. Therefore, the operators have to reallocate their virtualized resources accordingly. The operators may also decide only to resize the virtualized resources without altering the split. In this case, $o_n \!\neq\! 0 \!\!\iff\!\! \!(i_n \!\neq\! i_{n-1} \!\lor\! x_n \!\neq\! x_{n-1} \!\lor\! \hat{x}_n \!\neq\! \hat{x}_{n-1} )$. } activity ($o_n \neq 0$) is any change in the deployed split and virtualized resources (Figs. \ref{fig:migration1} and \ref{fig:migration2}). In this sense, deciding $o_n$ will directly determine split-$i$ as: $i_n \! := \! o_n \! \iff \! o_n  \! \neq \! 0$ and $i_n \! := \! i_{n-1} \! \iff \! o_n \!=\! 0$. Then, the allocated resources $x_n$ and $\hat{x}_n$ are determined through a mapping function $\omega \!:\! (\lambda_n,o_n) \!\mapsto\! (x_n, \hat{x}_n)$  (details in Sec. \ref{sec:algo}).

 In practice, $x_n$ and $\hat{x}_n$ may differ to the actual resource utilization.  We hence define $y_n \in \mathbb{R}$ and $\hat{y}_n \in \mathbb{R}$ as the actual resource utilization at vDU and vCU while deploying split $i_n \in \mathcal{I}$ and serving $\lambda_n$. The difference between the allocated resources and actual resource utilization is the prediction error resulting in overprovisioning or underprovisioning cost. If the operators do not want to reconfigure their resources, they may suffer from these costs, i.e., miss the opportunity to share unused resources for other workloads or even have declined demand due to insufficient capacity.
%
%
%
Otherwise, 
the operators can decide to resize their resources by $\beta_n \!\!:= \!\! |x_n \!-\! x_{n-1}|$ and $\hat{\beta}_n \!\! := \!\! |\hat{x}_{n} \!-\! \hat{x}_{n-1}|$ at vDU and vCU, respectively. 
Motivated by network slicing management \cite{aztec}, we propose a new cost model that can capture virtualized behaviour in vRANs. This model is illustrated in Fig. \ref{fig:orchestration} and described as follows.

\subsubsection{Overprovisioning} The operators have to pay more expensive cost for unused resources if the allocated resources are higher than the actual utilization. This cost is defined for resource overprovisioning, which is denoted:
\begin{align} \label{eq:overprov_cost}
	f_\text{o} \big( \max(0, x_n - y_n) + \max(0, \hat{x}_n - \hat{y}_n ) \big), 
\end{align}
%
where $f_\text{o}(.)$ is a function cost for overprovisioning. We assume this function is proportional with the input, e.g. $f_\text{o} (z) := \kappa_\text{o} z $, where $\kappa_\text{o}$ is the estimated fee for one unit capacity (\$/unit).


\subsubsection{Declined service demand} The underprovisioning and violating the split requirements can cause a disrupted or declined demand resulting in a penalty for service level agreement (SLA) violation and initializes a monetary compensation. We define this cost as:
\begin{align} \label{eq:underprov_cost}
	f_\text{d}
	\Big( \mathbbm{1}_{< y_n} (x_n) \lor \mathbbm{1}_{< \hat{y}_n} (\hat{x}_n) \lor \mathbbm{1}_{\neq 0} (C_n) \Big),
\end{align}
where $\mathbbm{1}(.)$ is an indicator function that takes value $1$ if the condition is satisfied, otherwise $0$. $C_n = (c_{np})_{\forall p}$ is a function of constraint dissatisfaction that captures the penalization for violating each $p$ constraints in vRANs at time $n$. The indicator functions in \eqref{eq:underprov_cost} are activated when the allocated resources $x_n$ and $\hat{x}_n$ do not meet the respective incurred resource utilization $y_n$ and $\hat{y}_n$. It is also activated when there exists constraint violations. In this case, the operator has to pay a penalty with a value $\kappa_\text{d}$ (\$) for any SLA violation. It complies with the monetary fee penalty used in network slicing \cite{deepcog}. 

\subsubsection{Instantiation \& Reconfiguration}
The operators may decide to reconfigure the split and virtualized resources following the users demand to reduce resource overprovisioning and avoid declined demand. However, there are overhead costs as:
%
%
\begin{align} \label{eq:reconfigure_cost}
	f_\text{i} \big( \beta_n   \mathbbm{1}_{> x_{n\!-\!1}} \! (x_n) \!+\!  \hat{\beta}_n  \mathbbm{1}_{> \hat{x}_{n\!-\!1}} \! (\hat{x}_{n}) \big) \!+\! f_\text{r} \big( (\beta_n  \!+\!  \hat{\beta}_n)  \mathbbm{1}_{\neq 0} ({o}_{n}) \big).
\end{align}
Instantiating and reconfiguring VMs have capital expenses in NFV \cite{aztec}. The first term in \eqref{eq:reconfigure_cost} captures the amount of instantiating additional resources at vDU and vCU. 
The second term then captures the cost initiated from any reconfiguration activities ($o_n \neq 0$). For instance, resizing the VMs' resources initiates a price of management delay \cite{5Gcoral} for load balancing setups and migrating the resources, which may interrupt the user sessions. We found that resizing a VM instance in CSC cPouta cloud (https://www.csc.fi/) induces delay around 23 seconds. Modern software architecture such as Kubernetes also needs several seconds for executing new pods \cite{aztec,5Gcoral}. The seamless migration for altering the split deployment also requires creating new vBS functions while preserving the old migrated functions active \cite{adaptive_vran_alba}.
Thus, the reconfiguration cost is affected by the migrated resources measured from the difference between the current and the previous virtualized resources.
 In our evaluation, we assume that both costs $f_\text{i}(.)$ and $f_\text{r}(.)$ are proportional to the input, e.g., $f_\text{i}(z) \!:=\! \kappa_\text{i} z$ and $f_\text{r}(z) \!:=\! \kappa_\text{r} z$, where $\kappa_\text{i}$ (\$/unit) is the parameter that captures the estimated cost for instantiation and $\kappa_\text{r}$ (\$/unit) is for reconfiguration. 

%
\textbf{xHaul cost.} O-RAN has proposed an open interface between vCUs, vDUs and RUs \cite{oran_architecture}. The operators can enjoy benefits such as computational and performance gains from a more centralized function \cite{complexity_cran}, but it incurs a higher transferred data load ($\delta$) \cite{vran_optimal_murti2}. S1 and S2 generate $\lambda$, S3 incurs $1.02 \lambda + 1.5$, while S4 transfers $2500$ (Mbps) of data load \cite{andres_fluidran_joint}. We define the xHaul cost as: $f_\text{h}(\delta_i) \!:=\! \kappa_\text{h} \delta_i$,  where $\delta_i$ is the data load for selecting split $i$ and $\kappa_{h}$ is the estimated fee for reserving a bandwidth (Mbps) of xHaul link.


%
\vspace{-2.25mm}
\subsection{Trade-offs and Problem Statement}
\vspace{-1mm}

\textbf{Trade-offs.} \textit{(i)} Centralizing more functions gains a lower computational cost, but it has tighter constraint requirements and requires a higher xHaul load. (\textit{ii)} STAO only incurs resource overprovisioning, but the amount of underutilized resources can be large. \textit{(iii)} DYNO can reduce the overprovisioning and insufficient allocated resources; however, it needs additional costs for reconfiguration and instantiation. \textit{(iv)} A standard loss function may not capture prediction error in virtualized resource management as the penalty fee is different for each behaviour. \textit{(v)} The relationship between the users' demand and the incurred virtualized resources is non-linear and non-monotonic. \textit{(vi)} A sequential decision for orchestrating the split selection and the virtualized resources at vCU and vDU is intricate problem, particularly without making assumptions of the underlying system and state transitions.
 
\textbf{Problem statement.} Given the above trade-offs, users' demand, and network state, what is the most suitable split and allocated resources at each time stage to minimize the long-term cost? 
%
%
%
Next, we present how we formulate the above sequential decision-making problem as an MDP, then discuss how LOFV solves the problem. 

\vspace{-2.5mm}
\section{Problem and Learning Framework} \label{sec:algo}
\vspace{-2mm}

\begin{figure}[t!]
	\centering
	\includegraphics[width=0.49 \textwidth]{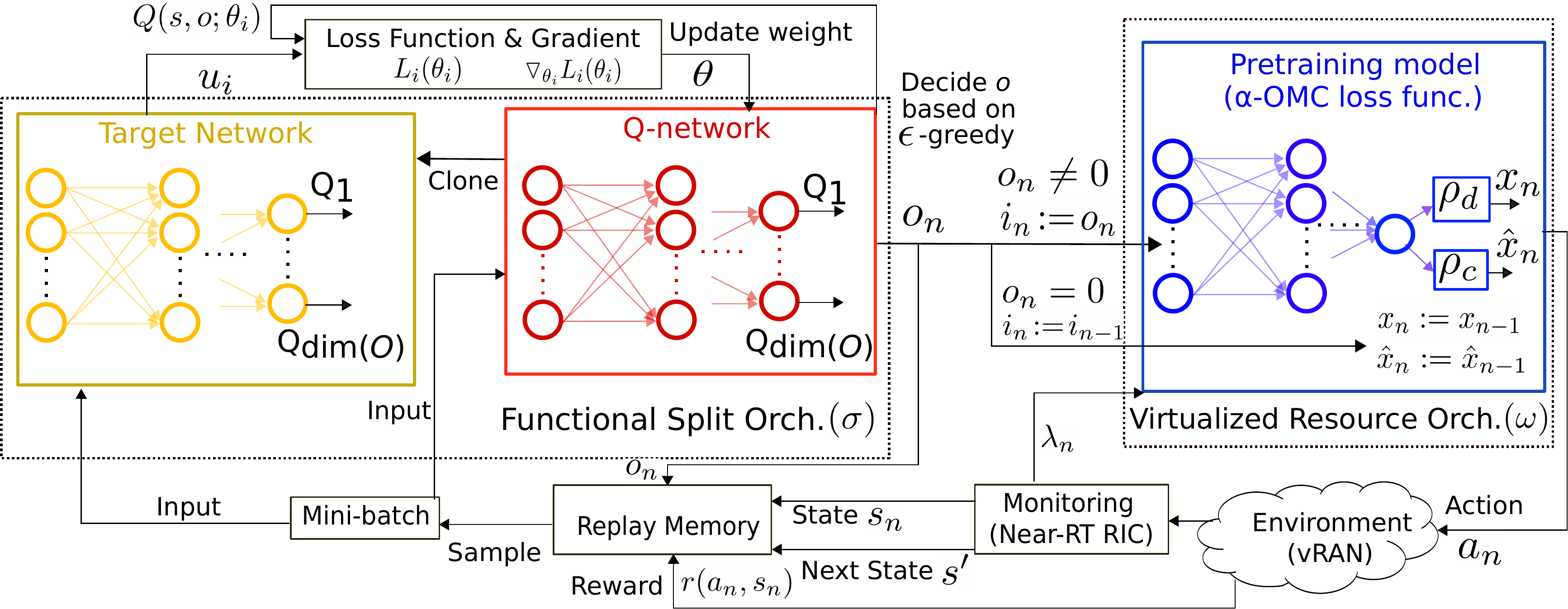}   
	\caption{\small \textbf{LOFV Architecture.} LOFV works following a model-free reinforcement learning and comprising of two functional blocks. 
	}
	\label{fig:drl_full}
	\vspace{-5.5mm}
\end{figure}
We formulate our problem in Sec. \ref{sec:model} as an MDP, which is specified by a tuple $\{ \mathcal{S}, \mathcal{A}, \mathcal{P}(s_{n+1} | s_n , a_n) , r_n, \gamma \}$. LOFV follows sequential decision-making based on a model-free reinforcement learning paradigm to solve the problem. At every time stage $n$, our agent observes a \textit{state} from state spaces ${s}_n \!\in\! \mathcal{S}$ drawn from the environment, takes an action ${a}_n$, and expects to receive a \textit{reward} signal $r({s}_n, {a}_n)$ as a feedback,  where $\gamma \in (0,1]$ is the discount factor. The state may not be stationary as the network load and conditions are changing over time with the sequence state arrival $({s}_n)_{n \in \mathcal{N}}$. Then, $\mathcal{P}(s_{n+1} | s_n, a_n)$ is the state transition probability that maps a state-action pair at time step $n$ into the distribution of next states. LOFV operation relies on two functional blocks: \textit{i) functional split orchestration ($\sigma$)}  and \textit{ii) virtualized resource orchestration ($\omega$)}. Fig. \ref{fig:drl_full} illustrates how LOFV operates.

\textbf{State.} 
Our state comprises: \textit{(i)} the incoming traffic demand at time stage $n$, $\lambda_n$; \textit{(ii)} the average traffic over $T$ period, $\bar{\lambda}_n \!:=\! \frac{1}{T} \sum_{t=1}^{T}\lambda_{nt}$; \textit{(iii)} the variance traffic for $T$ period, $\tilde{\lambda}_n \!:=\! \sum_{t=1}^{T} (\lambda_{nt} - \bar{\lambda})^2 / {T}$; \textit{(iv)} the previous allocated resources at vDU $x_{n-1}$ and \textit{(v)} vCU $\hat{x}_{n-1}$; and \textit{(vi)} the previous deployed split $i_n$. It provides \emph{time dynamic} of our variable interests: \textit{(i)} the input traffic demand that affects the split and resources; \textit{(ii-iii)} the characteristics and variation of the users demand' that helps to decide doing reconfiguration or not; and network conditions regarding \textit{(iv)} the availability of resources at vDU and \textit{(v)} vCU and \textit{(vi)} the last deployed split. Therefore, our agent receives an input state as a tuple ${s}_n = \{ \lambda_n, \bar{\lambda}_n, \tilde{\lambda}_n, x_{n-1}, \hat{x}_{n-1}, i_{n-1} \} \in \mathcal{S} \subseteq \mathbb{R}^{6} $.

\textbf{Action.} Our action space consists all pairs of $o \!\in\! \mathcal{O}$, $i \!\in\! \mathcal{I}$, $x \!\in\! \mathcal{X}$ and $\hat{x} \!\in\! \mathcal{\hat{X}}$. Thus, we define the action to be taken by our agent at the time stage $n$ as a tuple of ${a}_n \!=\! \{o_n, i_n, x_n, \hat{x}_n \} \!\in\! \mathcal{A}$, where $\mathcal{A} = \mathcal{O} \!\times\! \mathcal{I} \!\times\! \mathcal{X} \!\times\! \mathcal{\hat{X}}$ is our action spaces. 

\textbf{Reward.} Our objective is to minimize the long-term accumulated management cost over the time. Thereby, we define our reward function at time stage $n$ as: 
\begin{align} \label{eq:reward_function}
	r({a}_n, {s}_n) & := -J({a}_n, {s}_n)
\end{align}
where $ J(.):= \big(f_\text{o}(.) + f_\text{d}(.) + f_\text{i}(.) +  f_\text{r}(.) + f_\text{h}(.) \big)$ is the management cost defined in Sec. \ref{sec:model}.  The long-term accumulated reward starting at time step $n$ is $G_n \!:=\! \sum_{\tau=0}^{\infty} \gamma^\tau r_{\tau+n}$. Our goal can be redefined by maximizing the expected long-term accumulated reward as $\mathbb{E}[G_n] \!=\! \mathbb{E} [\sum_{\tau=0}^{\infty} \gamma^\tau r_{\tau+n}]$. Therefore, our agent aims to learn the optimal policy that maximizes the expected long-term accumulated reward as: $\pi_* := \arg \max \mathbb{E}_\pi [ \sum_{\tau=0}^\infty  \gamma^\tau r_{\tau+n}  | \pi]$,
%
%
%
%
%
where the policy $\pi$ is a function that maps from state to action $\pi({s}) \!:\! \mathcal{S} \!\mapsto\! \mathcal{A}$. To solve this maximization problem, we decompose our agent into two functional blocks and summarize its operation in Algorithm \ref{algo1}.

\begin{algorithm}[t!]  \caption{LOFV operation}
	\label{algo1}
	\SetAlgoLined
	\DontPrintSemicolon
	\textbf{Initialize:} Replay memory $\mathcal{D}$ with capacity $D$, Function $\omega_\varphi$ with pretaining weights $\varphi$, Q-network $Q_\theta$ with weight $\theta$,  Target Network $\hat{Q}_{\hat{\theta}}$ with weights $\hat{\theta} \leftarrow \theta$.  \\	
	\For{$e=1,..,E$ }{
		\textbf{Initialize:} $s_1 = \{ \lambda_{1}, 0, 0, x_{\text{max}}, \hat{x}_{\text{max}}, 1 \}$. \\
		\For{$n=1,..,N$}{
			Select random $o_n$ with probability $\epsilon$, otherwise $o_n := \max_o Q^*(s_n,o;\theta)$. \\
			Determine $i_n \! := \! o_n \! \iff \! o_n  \! \neq \! 0$ and $i_n \! := \! i_{n-1} \! \iff \! o_n \!=\! 0$ \\
			Allocate resources $(x_n, \hat{x}_n) \leftarrow \omega(\lambda_{n}, o_n)$  \\
			Execute $a_n = \{o_n, i_n, x_n, \hat{x}_n \}$ and observe reward $r_n$ \\
			Set $s_{n+1} := s_n$ \\
			Store transision $\big\{s_n, o_n, r_n, s_{n+1}  \big\}$ \\
			Sample random minibatch of transision $\big\{s_j, o_j, r_j, s_{j+1}  \big\}$ from $\mathcal{D}$. \\
			Set TD target $u_j := r_j + \gamma \max_{o'} \hat{Q}(s_{j+1}, o'; \hat{\theta})$ \\
			Perform a gradient descent on \eqref{eq:loss_q} with Adam
			\\
			$\hat{Q}_{\hat{\theta}} \leftarrow Q_\theta$ for every $C$ steps.
		}
	}
	\vspace*{-1.5mm}
\end{algorithm}
%
%

\vspace{-2mm}
\subsection{Functional Split Orchestration ($\sigma$) }
\vspace{-1mm}
In this block, we design a function $\sigma$ that maps the input state to the selected configuration setting $o$ and split $i$, thus the long-term accumulated reward is maximized. The reward function $r$ is affected by the configuration $o$, split-$i$ and virtualized resources at vDU $x$ and vCU $\hat{x}$. Given $o_n$, we can determine $i_n$ directly while allocate $x_n$ and $\hat{x}_n$ through a deterministic orchestrator $\omega$ (the virtualized resource orchestration). Consequently, in this block, we can treat $\omega$ as a part of the environment. We can redefine our action for this block with $o \in \mathcal{O}$ instead of using $ {a} \in \mathcal{A}$ which has a high dimensional action space. Then, the goal is to learn an optimal function that maximizes the long-term accumulated reward as: $\sigma_* := \arg \max_\sigma \sum_{\tau = 0}^{\infty} \gamma^\tau r_{n+\tau}.$
%
%
%
%
%
This problem is then has a high-dimensional state space but discrete action spaces. Hence, the function $\sigma$ can be designed using deep Q-learning \cite{dqn_mnih1}. It utilizes a neural network to approximate the action-value function (Q-function) while the learning step is based on Q-learning \cite{rl_sutton}.

We define the optimal action-value function as the maximum expected reward that follows some strategies based on a function $\sigma$, after observing some sequences ${s}$, then taking actions $o$ as: $Q^*({s}, o ) \!:=\! \max_\sigma \mathbb{E}[G_n | {s}_n \!=\! {s}, o_n \!=\! o]$. 
If the optimal value $Q^*({s}',o')$ of the sequence at the next time step ${s}'$ for all possible action $o'$ is known, we can define the optimal function $\sigma_*$, which is to select the action $o'$ that maximizes the expected value of $r + \gamma Q^*({s}', a')$ as: $Q^*({s}, o) \!:=\! \mathbb{E}_{{s} \sim \mathcal{E}} \big[ r + \gamma \max_{o'} Q^*({s}', o') \ | \ {s}', o' \big].$
%
%
%
%
%
%
Using \textit{value iteration}, the optimal action-value function can converge to the optimality when $i \!\rightarrow\! \infty$ \cite{rl_sutton}; however, this method is impractical \cite{dqn_mnih1}. Therefore, we use a function approximator to estimate the action-value function parameterized by weight $\theta$ as $ Q(\mathbf{s}, o; \theta) \!\approx\! Q(s,o)$. A neural network can be applied to approximate our action-value function (Q-network). The Q-network is trained by minimizing the sequence of a loss function that changes at iteration $i$:
\begin{align} \label{eq:loss_q}
	L_i(\theta_i) := \mathbb{E}_{{s}, o \sim p(.)} \big[ (u_i - Q({s},o; \theta_i) )^2 \big], 
\end{align}
where $u_i \!:=\! \mathbb{E}_{{s}' \sim \mathcal{S}} \big[ r + \gamma \max_{o'} \! Q(s',o'; \theta_{i-1} |  {s},o)\big]$ is the Temporal-Difference (TD) target and $p(s,a)$ is the \textit{behaviour distributions} (probability distribution over $s$ and $a$). Note that the parameters from the previous iteration $\theta_{i-1}$ is frozen while we are optimizing $L_i(\theta_i)$. 
%
%
%
%
%
Instead of directly calculating the full expectation of the gradient $\triangledown_{\theta_i} L_i(\theta_i)$, the loss function in \eqref{eq:loss_q} can be efficiently calculated using a stochastic gradient descent method, e.g., Adam optimizer \cite{adam_optim}. 
%
%
Then, our Q-network comprises of an input layer receiving $( \lambda_n, \bar{\lambda}_n, \tilde{\lambda}_n, x_{n-1}, \hat{x}_{n-1}, i_{n-1} )$, output layers activated by a linear function with a size of dim$({O})$, and the hidden layers constructed from a fully connected linear layer activated by ReLU function. This approach is a model-free and off-policy \cite{dqn_mnih1}. 
It also learns about the greedy strategy $o := \max_{o} Q(s,o; \theta$), while follows the behaviour distribution for exploration.  In this case, we can tailor an $\epsilon$-greedy strategy by setting the probability of selecting a random action with $\epsilon$; otherwise, we can select a greedy action by probability $1-\epsilon$. LOFV can directly determine split-$i$ from $o_n$ by $i_n \! \!:= \! o_n \! \iff \! o_n  \! \neq \! 0$ and $i_n \! := \! i_{n-1} \! \iff \! o_n \!=\! 0$.

\vspace{-2mm}
\subsection{Virtualized Resource Orchestration ($\omega$)}
\vspace{-1mm}
The virtualized resource orchestration is a function $\omega$ that maps an input pair of selected configuration $o$ (an output of a function $\sigma: s \mapsto o$) and the traffic demand $\lambda$ into the allocated virtualized resources at vDU $x$ and vCU $\hat{x}$. Thus, we define $\omega: (o,\lambda) \mapsto (x,\hat{x})$. This function has been simply defined as an affine function for vRAN network design \cite{andres_fluidran_joint}. However, our measurement shows that the relations between users' demand and virtualized resource utilization is non-linear and non-monotonic (Sec. \ref{sec:simulation}), which is hard to predict exactly. Therefore, we leverage a data-driven model that learns to approximate a function $\omega$. 
%
%
Our design is constructed from a simple neural network $\omega_\varphi$ parameterized by weight $\varphi$. The neural network consists of a single-neuron input layer receiving $\lambda$, a single-neuron output layer activated by a linear function and the hidden layers activated by a ReLU function.  The output of neural network $\omega_\varphi$ is then scaled with parameters $\rho_d$ (vDU) and $\rho_c$ (vCU) where each value depends on the split selection. The value of both parameters are obtained using a linear regression from \cite[Fig. 6a]{complexity_cran}. Then, we leverage $\alpha$-OMC \cite{deepcog} loss function to train $\omega_\varphi$ as a regressor, where $y$ (vDU) and $\hat{y}$ (vCU) are the ground truth. The $\alpha$-OMC is able to capture the prediction error with penalty fee incurred from resource overprovisioning and underprovisioning. As $\omega_\varphi$ is a deterministic orchestrator for $\sigma$, we can train separately and use it as a pretraining model. 
If $o_n=0$, our agent does not reallocate the resources, hence $x_n:= x_{n-1}$ and $\hat{x}_n:= \hat{x}_{n-1}$.

\vspace{-2mm}
\section{Results and Discussion} \label{sec:simulation}
%

\vspace{-1mm}
\subsection{Experimental Setup}
\vspace{-1mm}
Our testbed consists of the vBS, virtualized evolved packet core (vEPC) and virtualized user equipment (vUE), where each entity is built using an open-source srsRAN (e.g., srsENB, srsEPC, srsUE) \cite{srslte} and virtualized using a containerized system from Docker. The radio interface of vBS and vUE are emulated via radio samples in the Linux shared memory and inter-process communication. We deploy all of the virtualized entities in a general-purpose PC (AMD Ryzen 7 PRO 4750U) and use this computing specification as a \textit{reference core} (RC)\footnote{ We focus on the CPU resources as the most affected parameters; however, it is not limited and can be extended to other virtualized computing resources. We have 16 CPU threads in total, which translates to 100 RCs.}. We can control, set a capacity limit, and isolate each container resource through \textit{--cpus} in Docker.  The interval of time stage $t$ is a second and $n$ is a minute with default total duration per episode is 2 hours. Our default penalty coefficients are $\kappa_\text{o} \!\!=\!\! 1 $, $ \kappa_\text{d}\!\!=\!\! 2$, $\kappa_\text{i} \!\!=\!\! \kappa_\text{r} \!\!=\!\! 0.5$ and $\kappa_\text{h} \!\!=\!\! 0.0005$. 
The traffic demand follows a Poisson-generated user datagram protocol. 
The computing consumption of $f_1$, $f_2$ and $f_3$ yield $65\%$, $15\%$ and $20\%$ of total vBS computing usage, respectively \cite{andres_fluidran_joint}. The vCU consumes a twice of a lower computational processing than vDU (see  \cite[Fig. 6a]{complexity_cran} with $\approx$10 BSs). Thus, we have scaled parameters for deploying split-$i$ at vDU with $\rho^d_i \!\!=\!\! \{ 1, 0.8, 0.65, 0 \}$ and at vCU with $\rho^c_i \!\!=\!\! \{ 0, 0.1, 0.175, 0.5 \}$. 
The hyperparameters of our neural networks are set after exploratory assessments.
The neural network $\omega_\varphi$ is constructed from an input, an output and three hidden layers with the sizes of $128$, $64$ and $16$. We use Adam optimizer \cite{adam_optim} with learning rate is set to $5\!\times\! 10^{-5}$, mini-batch with the size of 128 and $\alpha$-OMC \cite{deepcog} loss function, then train $\omega_\varphi$ with 200 epochs. In the functional split orchestration, our Q-network consists of an input, outputs with the size of $|\mathcal{O}|$ and three hidden layers with the size of 512 each. The target network is updated every 10 time stage $n$. The batch size is set with 256 and the replay buffer has a capacity of $10^6$. Our exploration and exploitation strategy is based on $\epsilon$-greedy, where we set $\epsilon_{\max}=0.95$ at the beginning of episode, then it exponentially decays to $\epsilon_{\min} = 0.02$. We use Adam optimizer \cite{adam_optim} with learning rate is set to $0.0003$ and \eqref{eq:loss_q} for the loss function, then train our model over 5000 episodes.

%

\begin{figure}[t!]
	\centering
	\includegraphics[width=0.49 \textwidth]{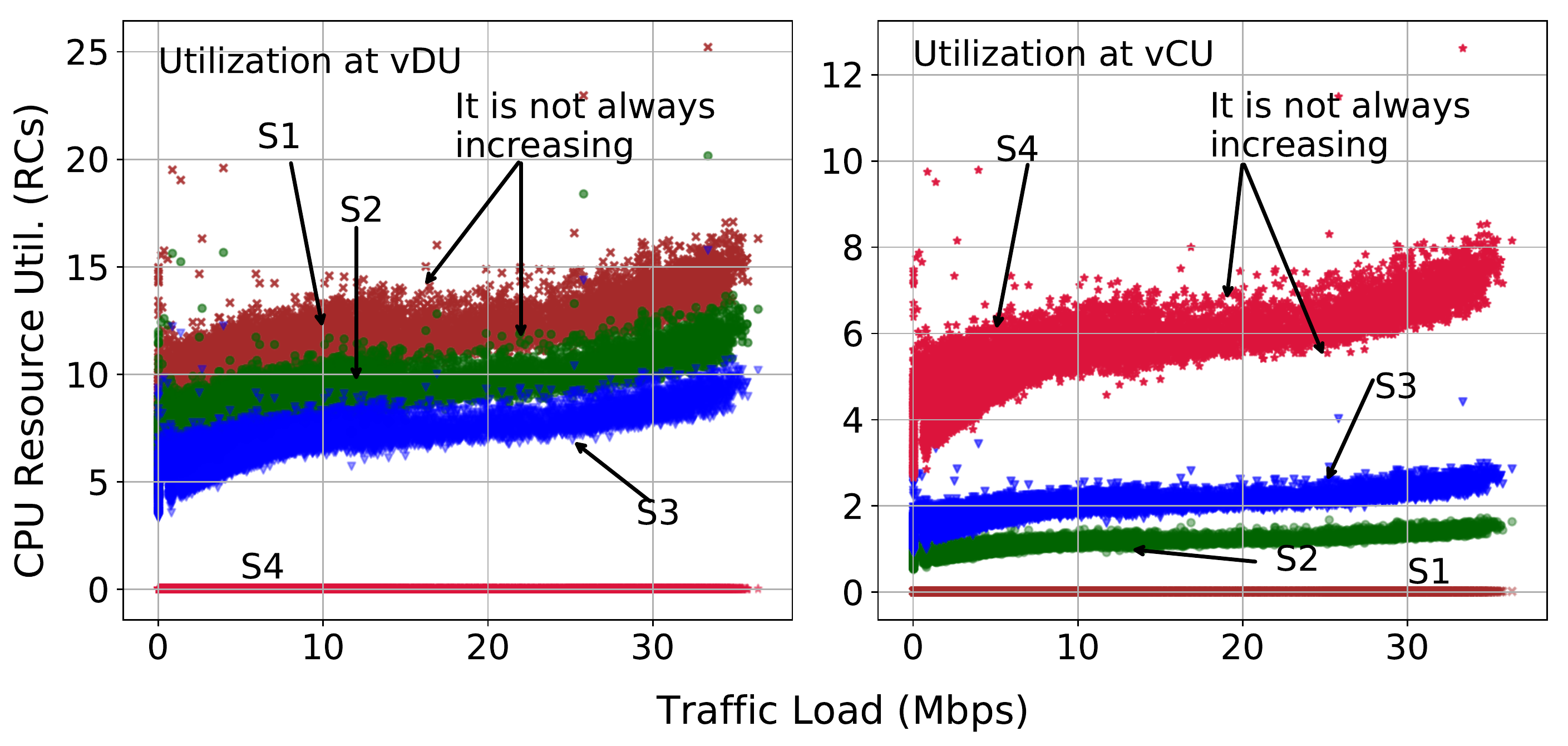} 
	\vspace{-4mm}  
	\caption{\small The relationship between the users' traffic demand to the incurred vBS resources at vDU (left) and vCU (right).}
	\label{fig:computing}
	\vspace{-6mm}
\end{figure}

\vspace{-2mm}
\subsection{Measurements and Evaluation Results}
\vspace{-1mm}

Our first evaluation studies the relationship between the users' demand and the virtualized resource utilization of vBS for selecting various splits. Fig. \ref{fig:computing} illustrates that the resource utilization at vDU and vCU for each split does not always increase with the growth of traffic load. Instead, it is non-linear and non-monotonic. Although the traffic load reaches a peak (35 Mbps), the virtualized resources are still highly underutilized with less than 25 RCs and 13 RCs of used resources at vDU and vCU, when LOFV is not implemented.


%
\begin{figure}[t!]
	\centering
	\includegraphics[width=0.275 \textwidth]{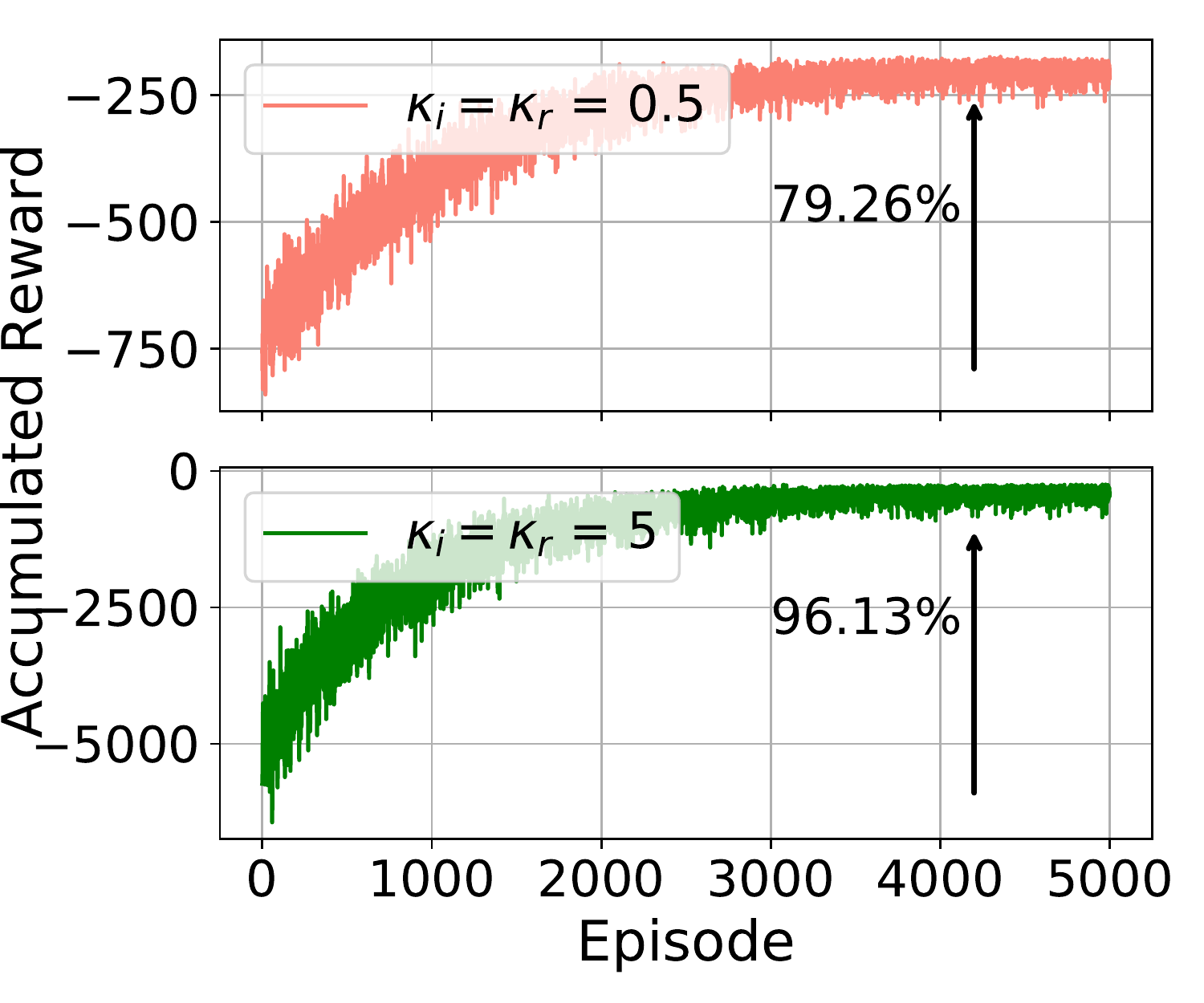}   
	\vspace{-2mm}
	\caption{\small Convergence performance of LOFV during training. \vspace{-2mm}}
	\label{fig:dqn}
	\vspace{-2mm}
\end{figure}
\begin{figure}[t!] 
	\centering
	\begin{subfigure}[c]{0.2355\textwidth}
		\includegraphics[width=\textwidth]{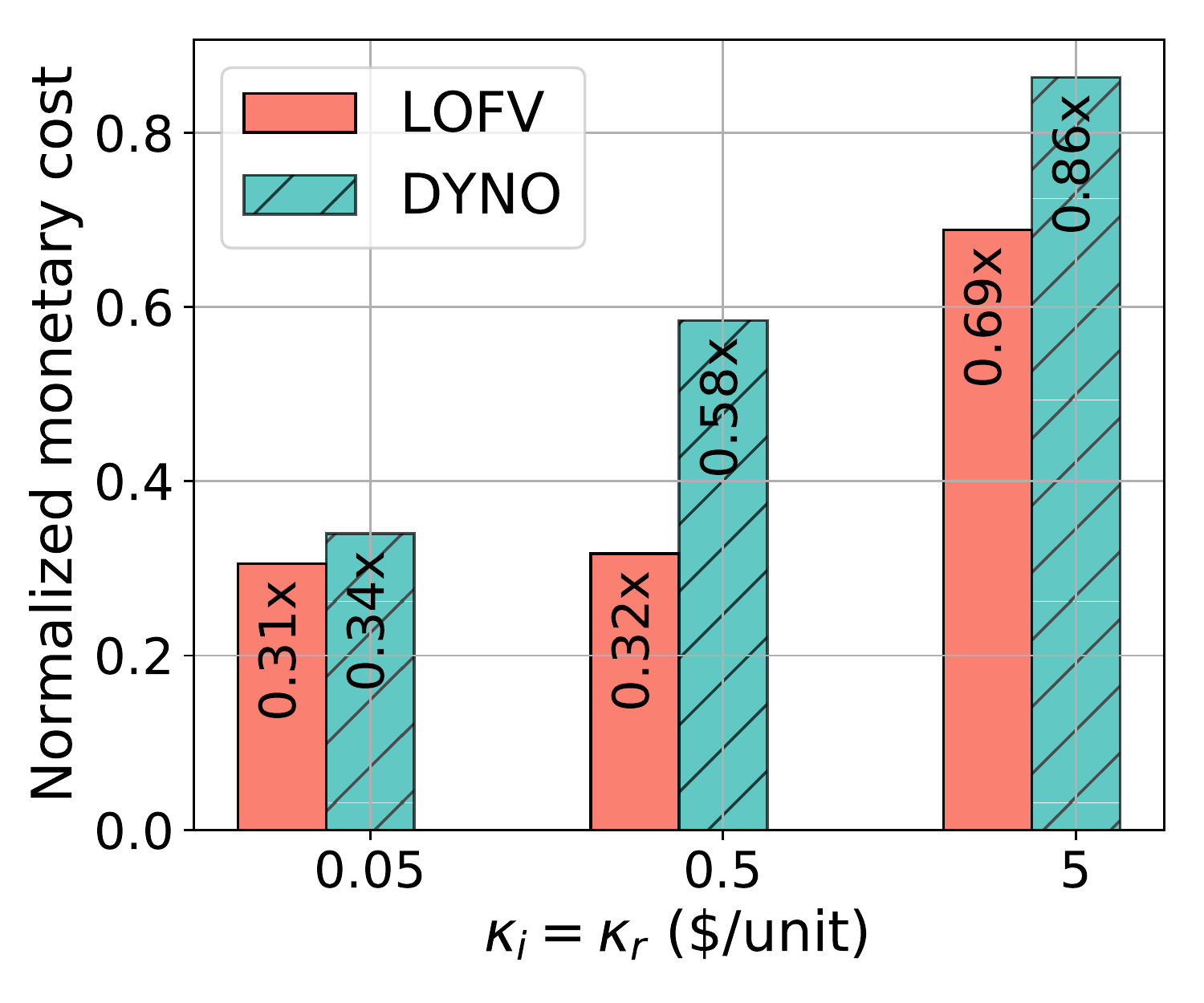}   
		\caption{}
		\label{fig:reconfig_c}
	\end{subfigure}
	\hfill
	\begin{subfigure}[c]{0.2355\textwidth}
		\includegraphics[width=\textwidth]{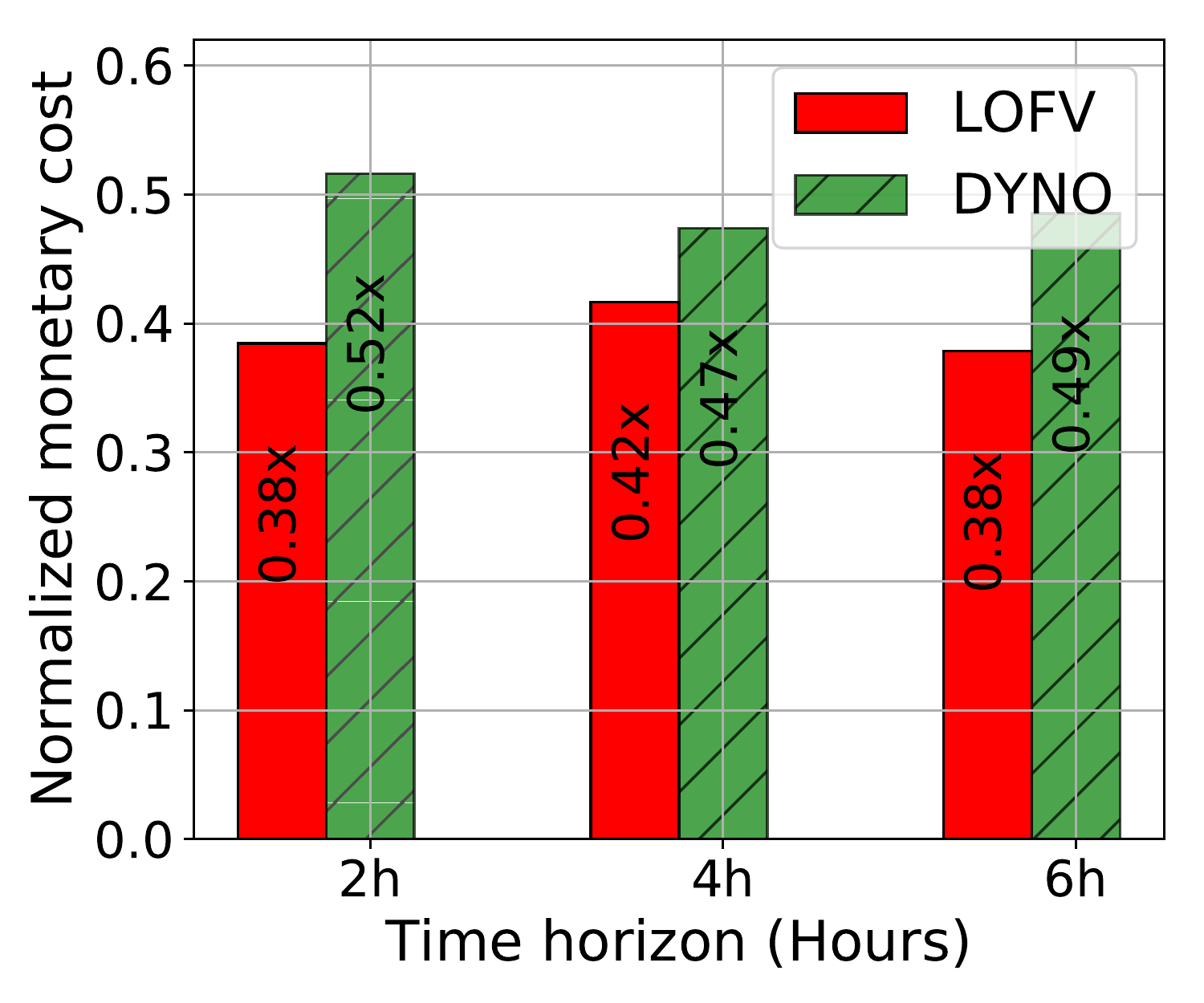}   
		\caption{}
		\label{fig:timehorizon_d}
	\end{subfigure}
	\vspace{-2mm}
	\caption{\small Performance of LOVF during the online stages with \textbf{(a)} various penalty coefficients and \textbf{(b)} over different time-horizon settings. All of the monetary costs are normalized toward STAO. }
	\vspace{-6mm}
\end{figure} 

Fig. \ref{fig:dqn} illustrates the convergence behaviour of LOFV during the training process. We increase the penalty coefficients of reconfiguration and instantiation from $\kappa_i \!\!=\!\! \kappa_r \!\!=\!\! 0.5$ (default) to $\kappa_i \!\!=\!\! \kappa_r \!\!=\!\! 5$. At the beginning of learning, LOFV has a higher probability from using a random policy for exploration, resulting in a high variance and expensive costs for both cases.  Also, a higher penalty coefficient ($\kappa_r=5$) induces a considerably more expensive cost (almost eight times) than a lower penalty ($\kappa_r \!\!=\!\! 0.5$) when choosing random actions. After some episodes, LOFV successfully learns the optimal policy and converges to the best policy that the agent can learn, and improves around 96.13\% ($\kappa_r \!\!=\!\! 5$) and 72.26\% ($\kappa_r \!\!=\! \!0.5$). 

We analyze the effect of penalty coefficients (reconfiguration \& instantiation) and the duration of time-horizon that LOFV can achieve during the online stages. Fig. \ref{fig:reconfig_c} depicts that LOFV has the lowest long-term cost across all coefficient settings compared to STAO and DYNO. It shows that LOFV achieves the highest cost-saving (69\%) compared to STAO when the penalty fees for reconfiguration and instantiation are low ($\kappa_r=0.05$). Although having a higher penalty coefficient ($\kappa_r=5$), LOFV still outperforms STAO with 31\% of cost-saving and has a cheaper cost compared to DYNO across all coefficients. LOFV manages to have 9\% of cost-saving at a low penalty coefficient ($\kappa_r\!\!=\!\!0.05$) and gains considerably cost-saving as high as 45\% of DYNO.

Fig. \ref{fig:timehorizon_d} shows the results of altering the duration of time-horizon (up to 6 hours) during the online stages that LOFV can achieve. It shows that the duration of the time horizon does not significantly degrade the performance of LOFV, where LOFV can achieve a stable cost-saving compared to our two baselines. LOFV saves around 58\% (4h) to 62\% (2h \& 6h) across all time-horizon settings compared to STAO. Similarly, LOVF also gains cost-saving to 27\% (2h) of DYNO.

\vspace{-2mm}
\section{Conclusion} \label{sec:conclusion}
\vspace{-1mm}
In this paper, we have proposed LOFV that jointly selects the splits and allocates the virtualized resources to minimize the long-term management cost. Aside from the xHaul cost, we have also proposed a new cost model accounting for resource overprovisioning, declined service demand, reconfiguration and instantiation, representing the NFV behaviour in vRANs. A model-free reinforcement learning paradigm has been leveraged to solve this sequential decision-making problem, constructed from deep Q-learning and a regression-based neural network. We also have performed a battery of tests to evaluate the effectiveness of LOFV using testbed measurements of a containerized RAN system. The results have shown that LOFV successfully learns the split selection and resource allocation and saves the cost by up to 69\% of STAO and 45\% of DYNO.



{
	\vspace{-2mm}
	\newcommand{\BIBdecl}{\setlength{\itemsep}{0.25 mm}}
	\bibliographystyle{IEEEtran}
	\bibliography{IEEEabrv,ref_lite}
}

\end{document}